\documentstyle[amssymb,amsmath,prd,aps,floats,epsfig,twoside]{revtex}
\def\bef{\begin{figure}}
\def\eef{\end{figure}}

\newcommand{\be}[1]{\begin{equation}\label{#1}}  
\newcommand{\beq}{\begin{equation}}
\newcommand{\eeq}{\end{equation}}
\def\eeq{\end{equation}}
\newcommand{\beqn}[1]{\begin{eqnarray}\label{#1}}
\newcommand{\eeqn}{\end{eqnarray}}
\newcommand{\bd}{\begin{displaymath}}
\newcommand{\ed}{\end{displaymath}}
\newcommand{\mat}[4]{\left(\begin{array}{cc}{#1}&{#2}\\{#3}&{#4}
\end{array}\right)}

\def\lsim{\raise0.3ex\hbox{$\;<$\kern-0.75em\raise-1.1ex
\hbox{$\sim\;$}}}
\def\gsim{\raise0.3ex\hbox{$\;>$\kern-0.75em\raise-1.1ex
\hbox{$\sim\;$}}} 
\def\simlt{\mathrel{\lower2.5pt\vbox{\lineskip=0pt\baselineskip=0pt
           \hbox{$<$}\hbox{$\sim$}}}}
\def\simgt{\mathrel{\lower2.5pt\vbox{\lineskip=0pt\baselineskip=0pt
           \hbox{$>$}\hbox{$\sim$}}}}
\def\unity{{\hbox{1\kern-.8mm l}}}

\def\lpr{l^\prime}
\def\phpr{\phi^\prime}
\def\hpr{h^\prime}
\def\Apr{A^\prime}
\def\hprdag{h^{\prime\dagger}}

\def\barl{\bar{l}} 
\def\barphi{\bar{\phi}} 
\def\barlpr{\bar{l}^\prime} 
\def\barphpr{\bar{\phi}^\prime} 

\renewcommand{\to}{\rightarrow}

\begin{document}
%
\renewcommand{\topfraction}{0.8}
\twocolumn[\hsize\textwidth\columnwidth\hsize\csname
@twocolumnfalse\endcsname

\title{ \vskip-1truecm{\hfill {\small CFNUL/01-04}} \\
\vskip0.001truecm{\hfill {\small DFAQ-2001/05-TH}} \\
Leptogenesis via Collisions: \\
Leaking Lepton Number to the Hidden Sector  }
\author{Lu{\'{\i}}s Bento$^a$ and Zurab Berezhiani$^{b,c}$ }
\address{
$^a$Centro de F\'{\i}sica Nuclear da Universidade de Lisboa, \\
Avenida Prof. Gama Pinto 2, 
1649-003 Lisboa, Portugal \\
$^b$Dipartamento di Fisica, Universit\'a di L'Aquila, I-67010 
Coppito, AQ,  Italy \\
INFN, Laboratori Nazionali del Gran Sasso, I-67010 Assergi, AQ, Italy \\
$^c$Andronikashvili Institute of Physics, GE-380077 Tbilisi, Georgia} 
\date{November, 2001}
\maketitle

\begin{abstract}
We propose a leptobaryogenesis mechanism   
in which the non-zero $B-L$ of the Universe is 
produced in out-of-equilibrium, lepton number and CP 
violating scattering processes that convert ordinary 
particles into particles of some hidden sector. 
In particular, we consider the processes 
$l\phi \to \lpr\phpr, \barlpr \barphpr$ mediated by the 
heavy Majorana neutrinos $N$ of the seesaw mechanism, 
where $l$ and $\phi$ are ordinary lepton and Higgs 
doublets and $\lpr$, $\phpr$ their hidden  
counterparts. Such a leptogenesis mechanism is effective  
even if the reheat temperature is much smaller than 
the heavy neutrino masses. In particular, it can be as low 
as $10^{9}$ GeV. 

\strut 

\end{abstract}

\strut

\newpage

]



It is well known that a non-zero baryon asymmetry (BA)    
can be produced in the initially baryon symmetric universe
if three conditions are fulfilled:
B-violation, CP-violation and departure from 
thermal equilibrium \cite{Sakh}.  These conditions 
can be satisfied in the decays of heavy particles 
of grand unified theories.  
On the other hand, the sphaleron processes,  
which violate $B+L$ but conserve $B-L$, 
are effective at temperatures from about $10^{12}$ GeV down to 
100 GeV \cite{KRS}. Thus, one actually needs to produce a 
non-zero $B-L$ rather than just $B$, a fact that disfavors 
the simplest baryogenesis picture based on grand unification 
models like $SU(5)$. When sphalerons are in equilibrium, 
the baryon number and  $B-L$ are related as $B = a(B-L)$, 
where $a$ is a model dependent order one coefficient \cite{bau}. 
Hence, in order to obtain the observed baryon to entropy density 
ratio $B=n_B/s = (0.6 - 1) \times
10^{-10}$, 
the produced $B-L$ needs to be ${\cal O}(10^{-10})$.  

The seesaw mechanism for neutrino masses offers 
an elegant possibility of generating non-zero $B-L$ 
in CP-violating decays of heavy Majorana neutrinos $N$ 
into leptons and Higgses,   
the so called leptogenesis scenario \cite{FY}.     
Namely, due to complex Yukawa constants, 
the decay rates 
$\Gamma (N\to l\phi)$ and $\Gamma (N\to \barl\barphi)$ 
can be different  from each other, 
so that leptons $l$ and anti-leptons $\barl$ 
are produced in different amounts.

In this Letter we propose an alternative mechanism 
of leptogenesis that 
is based on scattering processes rather than  particle decays. 
The main idea consists in the following. 
There may exist some hidden (shadow or mirror) sector of new 
particles which are not in thermal equilibrium with the 
ordinary particle world as far as the two systems interact very 
weakly e.g., if they only communicate via gravity.
However, other messengers may well exist namely, 
superheavy gauge singlets like right-handed neutrinos 
which can mediate very weak effective interactions 
between the ordinary  and shadow leptons. 
Then, a net $B-L$ could emerge in the Universe as a result of 
CP-violating effects in the unbalanced production of shadow 
particles from ordinary particle collisions.

The simplest model of this type can be described as follows. 
Consider the standard $SU(3)\times SU(2)\times U(1)$ model, 
containing three generations of leptons $l_i=(\nu, e)_i$, 
$e^c_i$ and quarks,
the Higgs doublet $\phi$, 
and some heavy singlet fermions $N_a$. 
Imagine now, that there is also a hidden sector with gauge symmetry 
$G^\prime$, containing fermion and scalar fields 
that are singlets under the standard model 
gauge group, while the ordinary particles  
are instead singlets under $G^\prime$.   
The interesting candidate can 
be a mirror sector, exact duplicate of the observable  
sector with the same gauge symmetry 
$G^\prime = SU(3)^\prime \times SU(2)^\prime \times U(1)^\prime$   
and with the same particle content \cite{LY,FV,BM,BDM,BCV}.  
However, in the more general case 
$G^\prime$ could be any gauge symmetry group containing, 
among other possible particles, fermions $\lpr_k$ 
and scalar $\phpr$ possessing opposite gauge charges so 
that the products $\lpr_k\phpr$ are gauge invariant.

In this case, the heavy singlet neutrinos $N$ can 
couple to $l,\phi$ as well as to $l^\prime,\phi^\prime$ 
and hence play the role of messengers between ordinary 
and shadow particles. 
The relevant Yukawa couplings have the form: 
\be{Yuk} 
h_{ia}l_i N_a \phi + \hpr_{ka}\lpr_k N_a \phpr + 
\frac{1}{2} M_{ab} N_a N_b  
+  {\rm H.C.}  
\end{equation} 
(charge-conjugation matrix $C$ is omitted); 
the $l,N,l^\prime$ states
are left handed while their $C$-conjugate, 
right-handed anti-particles are denoted as 
$\bar{l},\bar{N},\bar{l}^\prime$. 
Without loss of generality, 
the heavy neutrino mass matrix can be taken
in diagonal basis as $M_a = g_a M$, 
$M$ being the overall mass scale and $g_a$ order one 
real constants.  
After integrating out the heavy neutrinos $N$
the effective operators emerge 
from the couplings (\ref{Yuk}) as
\be{op} 
\frac{A_{ij}}{2 M} l_i l_j \phi \phi + 
\frac{D_{ik}}{M} l_i \lpr_k \phi \phpr + 
\frac{\Apr_{kn}}{2 M} \lpr_k \lpr_n \phpr \phpr  + {\rm H.C.} \;,
\end{equation}
with coupling constant matrices of the form 
$A = h g^{-1} h^T$, $\Apr = \hpr g^{-1} h^{\prime T}$  
and $D = h g^{-1} h^{\prime T}$.

Our mechanism works within the following scenario. 
We assume that the initial densities of the ordinary and 
hidden sectors are different from each other. 
In particular, 
the reheat temperature of the hidden sector 
should be smaller than the visible one,
$T'_R < T_R$, which can be achieved in certain 
inflationary models \cite{BDM,BCV,KST}.
The two particle systems interact very weakly
so that they do not come in  thermal equilibrium
with each other after reheating. 
The heavy neutrino masses are much larger than the 
reheat temperature $T_R$ and thus cannot be thermally 
produced. As a result, the usual 
leptogenesis mechanism via  $N\to l\phi$ decays is 
ineffective \cite{GPRT}. 
Now, the important role is played by 
lepton number violating scatterings 
mediated by the heavy neutrinos $N$ which stay out 
of equilibrium once $T_R \ll M$.  
On the other hand, they violate CP due to complex 
Yukawa couplings in Eq.\ (\ref{Yuk}).  

In other words, we assume that after the postinflationary  
reheating, different temperatures are established in the two 
sectors: the hidden world is cooler or ultimately, 
completely ``empty". 
Nevertheless, it starts to be ``slowly" 
occupied due to the leaking of entropy from the ordinary 
sector
through the reactions 
$l_i\phi \to \barlpr_k \barphpr$, 
$\barl_i\barphi \to \lpr_k \phpr$.   
Then, because  of  CP violation, 
the cross-sections with leptons and anti-leptons in the 
initial state 
are different from each other.
As a result, leptons leak to the hidden sector more (or less) 
effectively than antileptons 
and a non-zero $B-L$ is produced in the Universe.

A temperature scale that plays a crucial role in our 
considerations is the reheat temperature $T_R$,  
at which the inflaton decay and entropy production of the 
Universe is over and under which the Universe 
is dominated by a relativistic plasma of ordinary particle species.

It is convenient to introduce a parameter that characterizes 
the reaction rate per Hubble time at the temperature $T=T_R$, 
$K = (\Gamma/2H)_{R}$.  
Here $H = 1.66\, g_\ast^{1/2}T^2/M_{Pl}$ is the Hubble parameter 
and $g_\ast$ is the effective number of particle degrees 
of freedom.  
For the $\Delta L=1$ reaction rate we have 
$\Gamma_1 = \sigma_1 n_{eq} $, where 
$n_{eq}\simeq (1.2/\pi^2)T^3$ 
is an equilibrium density per degree of freedom and 
$\sigma_1$ is the total cross section of  
$l\phi \to \barlpr\barphpr$ scatterings, 
\begin{equation} \label{sigma} 
\sigma_{1} = \sum \sigma (l\phi \to \barlpr\barphpr) 
= \frac{Q_1}{8\pi M^2} \; . 
\end{equation}
The sum is taken over all flavor and isospin 
indices of initial and 
final states, and  $Q_1={\rm Tr}(D^\dagger D) = 
{\rm Tr}[(\hprdag\hpr) g^{-1}(h^\dagger h)^\ast g^{-1}]$. 
Hence, the out-of-equilibrium condition for this process reads as
\begin{equation} \label{K}
K_1 = \left(\frac{\Gamma_1}{2H}\right)_{R} \simeq 
1.5 \times
10^{-3}\, \frac{Q_1 T_R M_{Pl}}{g_\ast^{1/2}M^2} < 1 \,, 
\end{equation} 
which, for a given reheat temperature $T_R$, translates into 
the lower limit on the heavy neutrino mass scale $M$: 
\begin{equation}  \label{M1}
M_{12} > 1.3\, Q_1^{1/2} T_9^{1/2} , 
\end{equation}
where $M_{12}\equiv (M/10^{12}~{\rm GeV})$,  
$T_{9}\equiv (T_R/10^{9}~{\rm GeV})$ and 
we have taken $g_\ast\approx 100$ as in the standard model. 

However, there are also scattering processes like 
$l\phi \to \barl\barphi$ etc., which can wash out the 
produced $B-L$ unless they are out of 
equilibrium~\cite{FY,Luty92}. 
The total rate of $\Delta L=2$ processes 
is given by 
$\Gamma_2 \simeq (3 Q_2/4 \pi M^2)n_{\rm eq}$ 
where $Q_2 = {\rm Tr}(A^\dagger A) = 
{\rm Tr}[(h^\dagger h) g^{-1} (h^\dagger h)^\ast g^{-1}]$. 
Therefore, the condition $K_2=(\Gamma_2/2H)_{R} < 1$ 
translates into a lower bound similar to (\ref{M1}), 
\begin{equation}  \label{M2}
M_{12} > 3.2 \, Q_2^{1/2} T_9^{1/2} \, . 
\end{equation}
Clearly, if the Yukawa constants $h_{ia}$ and $\hpr_{ka}$ 
are of the same order, the out-of-equilibrium conditions 
for $\Delta L=1$ and $\Delta L=2$ processes 
are nearly equivalent to each other.

\begin{figure}[t]
\begin{center}
\epsfig{file=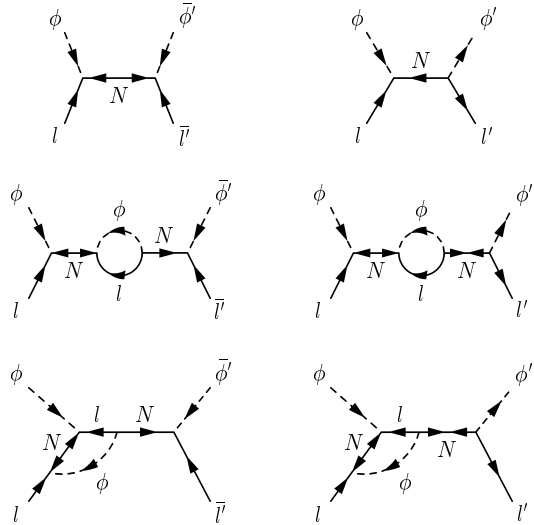,width=70mm} 
\end{center}
\caption{Tree-level and one-loop diagrams contributing to the 
CP-asymmetries in $l \phi \to \barlpr \barphpr$ 
(left column) and
$l \phi \to \lpr \phpr$ (right column).}
\label{fig1}
\end{figure}

\begin{figure}[t]
\begin{center}
\epsfig{file=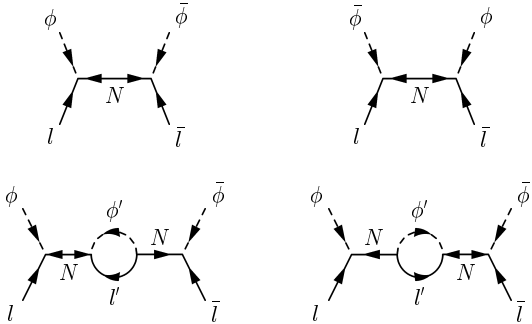,width=70mm}
\end{center}
\caption{Tree-level and one-loop diagrams contributing to the
CP-asymmetry of $l \phi \to \barl \barphi$.
The external leg labels identify the initial and final state particles.}
\label{fig2}
\end{figure}

Let us turn now to CP violation. 
In  $\Delta L=1$ processes the CP-odd lepton number
 asymmetry emerges from the
interference between the tree-level 
 and one-loop diagrams of Fig.\ \ref{fig1}. 
The tree-level amplitude for the dominant channel 
$l\phi\to \barlpr\barphpr$
goes as $ 1/M$
and the radiative corrections as $ 1/M^3$. 
For the channel $l\phi\to \lpr\phpr$ instead, 
both tree-level and one-loop amplitudes
go as $ 1/M^2$. 
As a result,  the cross section CP-asymmetries
are the same for both $l\phi\to \barlpr\barphpr$
and $l\phi\to \lpr\phpr$ channels
(on the contrary, the diagrams with $\lpr\phpr$ inside the loops, 
not shown in Fig.\ \ref{fig1}, 
yield asymmetries, $\pm \Delta \sigma'$, symmetric to each other). 
However, CP violation takes also place in $\Delta L=2$ 
processes (see Fig.\ \ref{fig2}). This is a consequence 
of the very existence of the hidden sector
namely, the contribution of the hidden particles to
 the one-loop diagrams of Fig.\ \ref{fig2}. 
The direct calculation gives: 
\begin{subequations}
\label{CP}
\begin{eqnarray}
&& 
\sigma (l\phi\to \barlpr\barphpr) -
\sigma(\barl\barphi \to \lpr\phpr) = 
(- \Delta\sigma  - \Delta\sigma' ) /2
\, ,  \\
&&  
\sigma (l\phi\to \lpr\phpr) -
\sigma(\barl\barphi \to \barlpr\barphpr) = 
( -\Delta\sigma + \Delta\sigma' )/2
\, ,   \\
&&  
\sigma (l\phi\to \barl\barphi) -
\sigma(\barl\barphi \to l\phi) = \Delta\sigma \, ;  \\
&& 
\Delta\sigma = \frac{3J\, S}{32\pi^2 M^4} \, ,  
\end{eqnarray}
\end{subequations}
where 
$J= {\rm Im\, Tr} [ (\hprdag\hpr) g^{-2}(h^\dagger h) g^{-1}
(h^\dagger h)^\ast g^{-1}]$  is the CP-violation parameter 
and $S$ is the c.m.\ energy square
($\Delta\sigma'$ is obtained from $\Delta\sigma$ 
by exchanging $h$ with $h'$).

This is in  perfect agreement with CPT invariance that 
requires that the total cross sections 
for particle and anti-particle scatterings are equal
to each other: 
$\sigma(l\phi \to X) = \sigma(\barl\barphi \to X)$. 
Indeed, taking also into account that by CPT, 
$\sigma(l\phi \to l\phi) = \sigma(\barl\barphi \to \barl\barphi)$,  
we obtain that the CP asymmetries in the $\Delta L=1$ and 
$\Delta L=2$ processes should be related as follows: 
\beqn{CPT}  
&&
\sigma(l\phi \to X^\prime) - \sigma(\barl\barphi \to X^\prime) +   
\nonumber \\ 
&&
\sigma (l\phi\to \barl\barphi) -
\sigma(\barl\barphi \to l\phi) = 0   \,,
\eeqn 
where $X^\prime$ are the hidden sector final states, 
$\barlpr\barphpr$ and $\lpr\phpr$.  
That is, the  $\Delta L=1$ and $\Delta L=2$ reactions have
CP asymmetries with equal intensities but opposite signs. 
But, as $L$ varies in each case by a different amount,
a net lepton number decrease is produced, or better,
 a net increase of $B-L$
$ \propto \Delta\sigma$.


Contrary to the lepton number that is violated by sphaleron processes,
$B-L$ is only violated by the above kind of reactions.
As long as we assume that the hidden sector is essentially depleted of
particles,
the only relevant reactions are the ones with ordinary particles in
the initial state.
Hence,  the evolution of the $B-L$ number density 
is determined by the CP 
asymmetries shown in Eqs.\ (\ref{CP}) and obeys the equation 
\begin{equation} \label{L-eq}
\frac{ d n_{B-L} }{dt} + 3H n_{B-L} = 
\frac34 \Delta\sigma \, n_{\rm eq}^2   \; .
\end{equation} 
Since the CP-asymmetric cross section 
$\Delta\sigma$ is proportional to the thermal average
c.m.\ energy square
$S \simeq 17\, T^2$ 
and  $H=1/2\, t \propto T^2$,
one integrates the above equation from $T=T_R$ to the 
low temperature limit and obtains 
the final $B-L$ asymmetry of the Universe as
\begin{equation} \label{BL} 
B-L = \frac{ n_{B-L} }{s} =
\left[\frac{\Delta\sigma\, n_{\rm eq}^2 }{4 H s} \right]_{R} \, ,
\end{equation} 
where $s$ is the entropy density.

The following remark is in order. 
In fact, the lepton number production starts as soon as the
inflaton starts to decay and the particle thermal bath is 
produced, before the reheat temperature is established. 
(Recall that the maximal temperature at the reheating period
 is usually 
larger than $T_R$.) In this epoch the Universe is still 
dominated by the inflaton oscillations and therefore it 
expands as $ t^{2/3}$ while the entropy of the Universe 
grows as $t^{5/4}$. The integration of Eq.\ (\ref{L-eq}) 
from some higher temperatures down to $T_R$ gives 
an asymmetry 1.5 times larger than the estimation (\ref{BL}). 
Taking all these into account, 
the final result can be recasted as follows: 
\begin{equation} \label{B-L}
B-L \approx 2 \times
10^{-3} \,\frac{J\, M_{Pl} T_R^3}
{g_\ast^{3/2} M^4  } \approx 
2 \times 10^{-8}\, \frac{J\, T_9^3}{M_{12}^{4}} \, ,
\end{equation}
where we have taken again $g_\ast \approx 100$. Taking also 
into account the lower limits (\ref{M1}) and (\ref{M2}), 
we obtain the upper limit on the produced $B-L$: 
\begin{equation} \label{upper}
B-L <   10^{-8}\, \frac{J\, T_9}{Q^2} \, ; \quad
Q = \max\{Q_1,  6\, Q_2 \}  \, .
\end{equation} 
This shows that for Yukawa constants spread e.g.\
in the range $0.1-1$ one can achieve $B-L= {\cal O}(10^{-10})$ 
for a reheat temperature as low as $T_R\sim 10^9$ GeV. 
Interestingly, this coincidence with the upper bound  
from the thermal gravitino production, 
$T_R < 4\times 10^9$ GeV or so \cite{R-therm}, 
indicates that our scenario could also work in the context 
of supersymmetric theories. Certainly, in a non-supersymmetric 
theory $T_R$  can be much larger.

The hidden sector may include coupling constants  
(e.g.\ gauge coupling constants of $G^\prime$) 
large enough to thermalize  
the hidden particles at a temperature $T^\prime$. 
Once $K_1 < 1$, $T^\prime$ will be smaller than the parallel 
temperature of the ordinary system $T$. 
Obviously, the presence of the out-of-equilibrium 
hidden sector does not affect much the Big Bang Nucleosynthesis
(BBN)  epoch. 
Indeed, if the two sectors do not come into full thermal 
equilibrium at temperatures $T\sim T_R$ then,
 they evolve independently during the Universe expansion
and approach the nucleosynthesis era with different 
temperatures.
For $K_1 < 1$, the energy density transferred to the hidden 
sector will be  crudely
$\rho^\prime \approx (8 K_1/g_\ast)\rho$, where 
$g_\ast(\approx 100)$ is attained to the leptogenesis 
epoch. Thus, assuming that at the BBN epoch the shadow 
sector is dominated by relativistic degrees of freedom, 
we obtain an effective number of extra light neutrinos
$\Delta N_\nu \approx K_1/2$.

Observe, that our model can induce the masses of both the 
ordinary and shadow neutrinos via their seesaw 
mixing with the heavy Majorana neutrinos 
provided that the shadow Higgs $\phi^\prime$ has a 
vacuum expectation value 
(VEV) $v^\prime \ll M$.  
The first operator in Eq.\ (\ref{op}), due to the ordinary 
Higgs VEV $\langle\phi^0 \rangle = v\sim 100$ GeV,  
induces the small Majorana masses of the ordinary neutrinos
while the other operators induce the mass and mixing mass terms
of the shadow neutrinos contained in $\lpr$ 
(in fact sterile neutrinos for the ordinary observer) 
with the ordinary active neutrinos~\cite{FV,BM}.
The total mass matrix of active-sterile neutrinos reads as~\cite{BM}   
\begin{equation} \label{numass} 
M_\nu = \mat{m_\nu}{m_{\nu\nu^\prime}}{m_{\nu\nu^\prime}^T}
{m_{\nu^\prime}} = 
\frac{1}{M} \mat{Av^2}{Dv v^\prime}{D^Tv v^\prime}
{A^\prime v^{\prime 2}} \, .
\end{equation}
In other words, it provides a simple explanation of
why sterile neutrinos could be light 
(on the same grounds as the active neutrinos) 
and could have significant mixing with the ordinary neutrinos. 
For example, if $v^\prime \sim 10^2 v$ then, the shadow 
neutrinos $\nu^\prime$ with masses of keV order could 
provide the warm dark matter component in the Universe 
\cite{BDM}. 
Instead, if $\langle\phi^\prime\rangle =0$, 
the $\nu^\prime$ are massless and unmixed 
with the ordinary neutrinos. 

It is worth noticing that the same mechanism that 
produces the lepton number in the ordinary Universe, 
can also produce the lepton prime asymmetry in 
the hidden sector. 
The amount of this asymmetry will 
depend on the CP-violation parameter that replaces
$J$ in Eqs.\ (\ref{CP}) and $\Delta \sigma'$ namely,
$J^\prime = {\rm Im\, Tr}[ (h^\dagger h) g^{-2}
(h^{\prime\dagger} h^\prime) g^{-1}
(h^{\prime\dagger} h^\prime)^\ast g^{-1}]$. 
Then, if the shadow sector contains also some 
heavier particles of the lepton or baryon type, 
the shadow matter could provide a dark matter. 
 
The interesting candidate is the mirror world, 
which has attracted a significant interest over the last years,
being motivated by various problems in particle 
physics and cosmology \cite{LY,FV,BM,BDM,BCV}.  
In this case we have a theory given by the
product $G\times G'$ of two identical gauge factors with
identical particle contents, which could naturally emerge
e.g.\ in the context of $E_8\times E'_8$ superstring theories.  
In particular, the $G$ sector contains ordinary particles 
$\phi$, $l$, etc., whereas $G^\prime$ contains their mirror partners 
$\phi^\prime$, $l^\prime$, etc., in equivalent 
representations.  
It is natural to assume that both particle sectors are described 
by identical Lagrangians, 
that is, all coupling constants (gauge, Yukawa, Higgs)
have the same pattern in both sectors and thus
their microphysics is the same.

In particular, a discrete symmetry under the exchange 
$\phi \to \phi^{\prime\dagger}$, $l \to \bar{l}^\prime$, etc.,
the so-called mirror parity, 
implies $h^\prime_{ia} = h^\ast_{ia}$. 
In this case the CP-violation parameters are the same,
$J^\prime = J$.  
Then, 
one expects that $n_{B-L} = n^\prime_{B-L}$ and the 
mirror baryon number density should be equal to the ordinary 
baryon density, $\Omega_B^\prime = \Omega_B$. 
The mirror parity could be also spontaneously broken 
by the difference in weak scales
$\langle \phi \rangle =v$ and $\langle \phi' \rangle =v'$, 
which would lead to somewhat different particle
physics in the mirror sector \cite{BM,BDM}, 
e.g.\ the mirror leptons and baryons could be heavier 
(or lighter) than the ordinary ones. 
But, as the mechanism only depends 
on the Yukawa constant pattern in (\ref{Yuk}),
one still has $n_{B-L} = n^\prime_{B-L}$, while 
$\Omega_B^\prime \neq \Omega_B$. 
Generically, the mirror sector provides a sort of 
self-interacting dark matter, however, if it is
significantly colder than the visible one, 
$T^\prime/T < 0.3$ or so, 
the mirror photons decouple early and the mirror 
matter would behave as a cold dark matter 
as far as the large scale formation is concerned \cite{BCV}. 
Moreover, the mirror group $SU(2)'\times U(1)'$ may be even
fully broken by a set of two or more Higgs VEVs,
which would make the mirror photon a massive particle
and thus the corresponding interaction short-range.

It is worthwhile to stress that the leptogenesis mechanism  
we propose does not really rely on  model dependent features 
of the hidden sector.
They are however important and 
in principle testable to some extent
if the hidden sector is to make up for the dark matter 
of the Universe. Depending on the gauge structure, 
field content and symmetry breaking scales 
in the hidden sector, one could have a shadow matter 
behaving as a cold, warm or self-interacting dark matter, 
or their combination.  
The possible marriage between dark matter and the
leptobaryogenesis mechanism  
is certainly an atractive feature of our scheme  
which deserves to be explored in more detail.

Let us conclude with the following remark. 
The magnitude of the produced $B-L$, Eq.\ (\ref{B-L}), 
strongly depends on the temperature -- 
namely, larger $B-L$ will be produced in 
the patches where the plasma is hotter. 
In the cosmological context, this would lead to 
a situation where apart from the adiabatic density/temperature 
perturbations, there also emerge correlated isocurvature 
fluctuations with variable $B$ and $L$ which could be tested 
with the future data on the CMB anisotropies and large scale 
structure.  

\vskip 5pt

We thank A. Dolgov and E. Paschos for discussions. 
We acknowledge  Funda\c c\~ao para a  Ci\^encia e Tecnologia (FCT) 
for grant CERN/P/FIS/40129/2000.  
The work of Z. B. was partially supported by the MURST research 
grant {\it ``Astroparticle Physics"}. 

\vspace{-15pt}


\end{document}